\documentclass[sigconf]{acmart}

\AtBeginDocument{%
  }

\copyrightyear{2024}
\acmYear{2024}
\setcopyright{rightsretained}
\acmConference[PACT '24]{International Conference on Parallel Architectures and Compilation Techniques}{October 14--16, 2024}{Long Beach, CA, USA}
\acmBooktitle{International Conference on Parallel Architectures and Compilation Techniques (PACT '24), October 14--16, 2024, Long Beach, CA, USA}
\acmDOI{10.1145/3656019.3689612}
\acmISBN{979-8-4007-0631-8/24/10}

\usepackage{pifont}
\usepackage{xspace}
\usepackage{multirow}
\usepackage{makecell}
\usepackage[per-mode=symbol, range-units=single, mode=text]{siunitx}
\newcommand{\kB}{\kilo B}
\newcommand{\MB}{\mega B}
\newcommand{\GB}{\giga B}
\newcommand{\TB}{\tera B}

\definecolor{darkorchid}{rgb}{0.6, 0.2, 0.8}
\usepackage[capitalise, nameinlink, noabbrev]{cleveref}

\newcommand{\figvspace}[1]{\vspace{#1}}
\newcommand{\tablefontsize}{\footnotesize}

\newcommand{\design}{Trimma\xspace}
\newcommand{\irt}{iRT\xspace}
\newcommand{\irc}{iRC\xspace}

\begin{document}

\title{\design: Trimming Metadata Storage and Latency for Hybrid Memory Systems}


\author{Yiwei Li}
\affiliation{%
  \institution{Tsinghua University}
  \city{Beijing}
  \country{China}}
\email{liyw19@mails.tsinghua.edu.cn}

\author{Boyu Tian}
\affiliation{%
  \institution{Tsinghua University}
  \city{Beijing}
  \country{China}}
\email{tby20@mails.tsinghua.edu.cn}
\author{Mingyu Gao}
\affiliation{%
  \institution{Tsinghua University}
  \city{Beijing}
  \country{China}}
\affiliation{%
  \institution{Shanghai Qi Zhi Institute}
  \city{Shanghai}
  \country{China}}
\email{gaomy@tsinghua.edu.cn}

\begin{abstract}
Hybrid main memory systems combine both performance and capacity advantages from heterogeneous memory technologies. With larger capacities, higher associativities, and finer granularities, hybrid memory systems currently exhibit significant metadata storage and lookup overheads for flexibly remapping data blocks between the two memory tiers. To alleviate the inefficiencies of existing designs, we propose \design, the combination of a multi-level metadata structure and an efficient metadata cache design. \design uses a multi-level metadata table to only track truly necessary address remap entries. The saved memory space is effectively utilized as extra DRAM cache capacity to improve performance. \design also uses separate formats to store the entries with non-identity and identity address mappings. This improves the overall remap cache hit rate, further boosting the performance. \design is transparent to software and compatible with various types of hybrid memory systems. When evaluated on a representative hybrid memory system with HBM3 and DDR5, \design achieves up to 1.68$\times$ and on average 1.33$\times$ speedup benefits, compared to state-of-the-art hybrid memory designs. These results show that \design effectively addresses metadata management overheads, especially for future scalable large-scale hybrid memory architectures.

\end{abstract}




\maketitle

\section{Introduction}\label{sec:introduction}

Main memory now performs an increasingly critical role in computer systems, especially when executing data-intensive applications with massive amounts of data~\cite{mckee2004reflections}. Both fast access speed and large capacity are desired for main memory, but these two goals exhibit fundamental conflicts under any existing individual memory technology. Fortunately, hybrid memory systems~\cite{lh-cache, alloycache, unisoncache, footprintcache, footprint_tagless_cache, banshee, batman, pageplacement-ics11, hma15, pom14, cameo, silc-fm, mempod}, which integrate two or more memory tiers with different characteristics, have the potential to overcome such a tradeoff and achieve both performance and capacity advantages over traditional architectures. Typically, a slow memory tier provides a sufficiently large capacity to hold the entire dataset, while a fast memory tier delivers lower latency and higher bandwidth for the most frequently accessed data.

With new technologies such as non-volatile memories (NVMs)~\cite{pcm, optanedcpmm} and new interconnection techniques such as Compute Express Link (CXL)~\cite{cxl}, the slow memory capacity is able to keep increasing, while the fast memory chooses to maintain its small size for sufficiently fast access speed. Furthermore, to improve performance, hybrid memory systems start to adopt higher associativities for address mapping between the two tiers~\cite{tagless_cache, lgm, mempod, hybrid2}, and also use more fine-grained data blocks for migration~\cite{cameo, silc-fm, hybrid2, lh-cache, footprintcache}. These trends inevitably require highly flexible address remapping schemes covering the entire memory space, so that the most critical data can be effectively identified and moved to the fast memory.

Consequently, in future hybrid memory systems, not only do we need to \emph{store} a large amount of metadata for address remapping, but also these metadata must support sufficiently fast \emph{lookup} to avoid being a performance bottleneck. 
Unfortunately, existing metadata designs fail to alleviate these storage and lookup overheads when hybrid memory systems scale to large capacities, high associativities, and fine-grained data blocks. 
For example, cache-style tag matching schemes~\cite{lh-cache, silc-fm, banshee, alloycache, unisoncache} store address tags only for data blocks in the fast memory, thus exhibiting moderate metadata storage cost. However, associative tag matching cannot support high-associativity designs due to the significant latency to search through the many tags inside a cache set.
Alternatively, the address remapping information can be kept using a simple linear table~\cite{pom14, mempod, chameleon, hybrid2}, where metadata access only requires a single lookup. Dedicated on-chip caches can further alleviate this access latency~\cite{pom14, chameleon, mempod, hybrid2, atcache}. But on the other hand, because the remap table needs to have an entry for \emph{every} data block, it has a huge size that proportionally grows with the \emph{total} memory capacity. Such a huge table needs to be stored in the fast memory, occupying the available high-performance capacity for the actual data. Moreover, the SRAM metadata cache also incurs on-chip area cost and may not scale to effectively buffer the increasingly larger metadata.

To alleviate such metadata overheads, in this work we propose \design, the combination of a multi-level metadata structure and an efficient metadata cache design for hybrid memory systems. \design is compatible with both cache and flat use modes of the fast memory. It is a hardware-only design that is completely transparent to application and system software. These features make \design an easy-to-adopt and widely applicable architecture.

To reduce the metadata \emph{storage} cost, \design uses an \emph{indirection-based remap table} (\irt), similar to the multi-level OS page table, to only store the truly necessary metadata. \irt is a scalable structure as its size is proportional to only the fast memory capacity instead of the overall memory size under most cases. To achieve this, \irt eliminates the entries for unallocated \emph{and uncached/unmigrated} data blocks, whose mappings are always the identity function. With a simple yet efficient memory management scheme, \irt can easily allocate/deallocate entries in hardware. Despite being constantly varying and highly fragmented with irregular distribution, the saved fast memory spaces for the unnecessary metadata entries are effectively utilized by \design as extra DRAM cache slots to deliver performance gains to the system. 

In addition, to reduce the metadata \emph{lookup} latency, especially the even more expensive multi-level traversals in \irt, \design uses an \emph{identity-mapping-aware remap cache} (\irc), with separate formats to store the entries with non-identity and identity mappings. The non-identity mappings use the same conventional remap cache design, but with a slightly smaller capacity to save some space for storing identity mappings. The identity-mapping part of \irc instead uses a different and more compact format to store more entries within the same amount of SRAM. This is because we only need to know which addresses exhibit identity mappings, without the need to store the remapped pointers. 

With these two key techniques, \design is able to offer significant performance improvements. 
We compare \design to state-of-the-art designs under different hybrid memory technology combinations. 
\design achieves up to 1.68$\times$ and on average 1.33$\times$ speedups on an HBM3 + DDR5 system, and up to 1.80$\times$ and on average 1.34$\times$ speedups on a DDR5 + NVM system.
The reasons for such improvements are twofold. First, there is a large extra DRAM cache area due to metadata savings, from uncached/unmigrated data with identity address mappings. Data accesses missed from the fast memory in \design are therefore reduced by 7.9\% and migration traffic is reduced by 23\%.
Second, the identity-mapping-aware remap cache effectively increases the coverage of the remap cache, and reduces metadata access misses. The overall remap cache hit rate has increased from 54\% to 67\%. 
Overall, our findings indicate that \design effectively improves the scalability of hybrid memory systems by tackling the challenges associated with metadata storage and lookup overheads, particularly suitable for upcoming large-scale hybrid memory architectures.

In summary, we make the following contributions in this paper.
\begin{itemize}
    \item We identify that the metadata storage and lookup overheads would become potential bottlenecks in hybrid memory systems, given the recent trends towards larger capacities, higher associativities, and more fine-grained block sizes.
    \item We propose an indirection-based remap table structure, \irt, that only keeps truly necessary remap entries. The memory space for unnecessary remap entries is retrofitted as extra DRAM caching space, hence improving performance.
    \item We propose an efficient identity-mapping-aware remap cache design, \irc, that uses separate formats and storage for non-identity and identity address mappings. Particularly, the identity mappings use a more compact scheme to improve the remap cache coverage and thus the hit rate.
    \item We integrate the two techniques in \design, a scalable hybrid memory metadata management scheme. Our evaluation shows that \design achieves significant performance benefits on various hybrid memory systems over the state-of-the-art designs.
\end{itemize}


\section{Background and Motivations}\label{sec:motivation}

Hybrid memory systems typically have multiple memory tiers, either by integrating two or more heterogeneous memory technologies, such as 3D-stacked DRAM~\cite{hmc, hbm, hbm3}, DDR4/5 DRAM~\cite{ddr4, ddr5}, and/or byte-addressable non-volatile memories (NVMs)~\cite{optanedcpmm}, or extending beyond local memory to remote memory through advanced interconnection technologies like Compute Express Link (CXL)~\cite{cxl}. 
The \emph{fast memory} tier usually has lower access latency and higher bandwidth, but comes with smaller capacity. In contrast, the \emph{slow memory} tier offers much larger capacity to extend the system main memory at lower cost, while its access speed is inferior. 
Hybrid memory systems can either use the fast memory as a cache in front of the slow memory~\cite{lh-cache, alloycache, unisoncache, footprintcache, footprint_tagless_cache, banshee, batman, accord}, or treat the two as a horizontally flat organization both visible to the OS~\cite{pageplacement-ics11, hma15, pom14, cameo, silc-fm, mempod}. Data are cached or migrated between the two tiers to let the most critical subset reside in the fast memory. A large volume of recent research has been focusing on specific cache management and flat migration policies~\cite{gran_aware_migration, banshee, batman, hma15, thermostat, pageplacement-ics11, bw_aware_page_placement, bear}, in order to exploit the technology heterogeneity and achieve both performance and capacity advantages over traditional architectures.

\subsection{Trends of Hybrid Memory Systems}\label{sec:motivation:trends}

With technology advances and system development, hybrid memory systems are currently observing several architectural trends. First, the typical memory capacities of the fast and slow memories keep diverging, resulting in \textbf{increasingly higher slow-to-fast capacity ratios}. 
Memory-intensive big-data applications like neural networks~\cite{gradient, imagenet, deep-learning, video-cnn}, genome alignment~\cite{altschul1990basic, harris2007improved, graph-genome}, and graph analytics~\cite{ligra, lightweight-graph, graphmat, graph-for-human, ringo} have been continuously driving the growth of main memory capacity. 
The emerging NVM technologies, such as phase-change memories~\cite{pcm} and Intel 3D XPoint~\cite{optanedcpmm}, promise to handily extend the byte-addressable memory in a system to \si{\TB} scales. Similarly, advanced interconnection technologies like CXL 3.0 switches could also extend the system's remote memory capacity to several \si{\TB}s~\cite{cxl3-whitepaper, micron_cz120}. 
In contrast, fast memory technologies, primarily the in-package 3D-stacked HBM modules, are limited in their capacity due to physical area cost, vertical integration difficulty, and thermal dissipation requirements. Currently even the most recent HBM3 only offers a maximum of \SI{24}{\GB} per stack~\cite{hbm3}, two orders of magnitude smaller than an NVM-based slow memory.
As a concrete example, the Intel Sapphire Rapids processors~\cite{intel_spr} integrate four HBM2E stacks and eight DDR5 channels, theoretically capable to scale to \SI{64}{\GB} HBM and \SI{4}{\TB} DRAM, resulting in a slow-to-fast capacity ratio as high as 64:1. 

Second, the increasing slow-to-fast capacity ratio necessitates hybrid memory systems to adopt \textbf{higher associativities for mapping between fast and slow memories}. Early hybrid memory systems use very low associativities, usually partitioning the two memories into many small sets with only one fast block per set which several slow blocks can map to~\cite{pom14, cameo, chameleon, alloycache}. With increasing capacity divergence, many slow blocks will compete for the single slot in the fast memory, incurring significant conflict misses and frequent replacements. To relieve the rigid mapping bottleneck, systems with higher associativities are recently proposed~\cite{tagless_cache, lgm, mempod, hybrid2}, and result in higher hit rates in the fast memory.
As shown in \cref{fig:motiv:metadata}, we run \texttt{PageRank} in a 16-core system with hybrid memory of DDR5 and HBM3, under a capacity ratio of 32:1 (detailed configurations are in \cref{sec:methodology}).
If we do not consider the metadata overheads (discussed in \cref{sec:motivation:challenges}), increasing the associativity from 1 to 1024 could bring a significant speedup of 1.5$\times$ in the ideal case.

Third, hybrid memory systems would prefer \textbf{fine-grained blocks}. Previous designs have demonstrated the tradeoff regarding data block granularities~\cite{pom14, mempod, hybrid2, tagtables, gran_aware_migration}. In general, coarse-grained blocks exploit spatial locality, but also result in over-fetching that wastes significant memory bandwidth. A larger granularity also leads to fewer blocks under a fixed capacity, and incurs more capacity and conflict misses. On the other hand, fine-grained blocks increase both capacity and bandwidth utilization, but the large number of blocks may result in substantial management overheads such as metadata storage. Several designs have used sub-blocking techniques to balance bandwidth utilization and metadata overheads~\cite{footprintcache, silc-fm, sectorcache, hybrid2}, i.e., use a coarse-grained block size but only fetch the demanded fine-grained sub-blocks. In summary, most hardware hybrid memory systems prefer block granularities smaller than the OS page size, from a few hundreds of bytes (\SI{64}{B} or \SI{256}{B})~\cite{cameo, silc-fm, hybrid2, lh-cache, footprintcache} to the DRAM page size (\SI{2}{\kB})~\cite{pom14, hybrid2}.

\subsection{Challenges of Metadata}\label{sec:motivation:challenges}

In this work, we focus on hardware-only hybrid memory management that is completely transparent to software. Because any data block could potentially be moved across memory tiers, \emph{every} data access request to a \emph{physical address} must first lookup a hardware-based \emph{metadata} structure to determine the actual location of this data block, i.e., the \emph{device address} on the fast or slow memory tiers. As a result, the metadata design is particularly vital in terms of both performance and system cost. 
We discuss several existing metadata schemes below, and show how they would become a new potential bottleneck, especially in future hybrid memory systems that are scaling to large slow-to-fast capacity ratios, high associativities, and fine block granularities, following the trends in \cref{sec:motivation:trends}.

\begin{figure}
\centering
\includegraphics[width=\columnwidth]{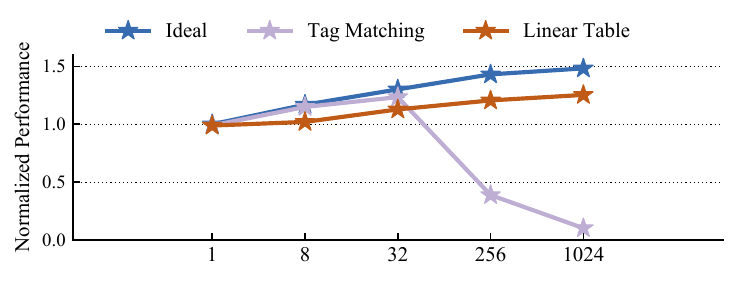}
\caption{Performance comparison among various metadata management schemes for the \texttt{PageRank} workload. Simulation details are in \cref{sec:methodology}. ``Ideal'' represents the theoretical scenario without metadata storage and lookup overheads. Normalized to the ideal case at an associativity of 1.}
\label{fig:motiv:metadata}
\end{figure}

One popular metadata structure is the cache-style \emph{tag matching} scheme, used by many prior systems, especially DRAM cache designs~\cite{lh-cache, silc-fm, bimodal-cache, bear, banshee, alloycache, unisoncache}. Within a specific cache set in the fast memory (DRAM cache), all tags must be compared against the access request. Upon finding a matched tag, the corresponding fast memory address is the remapped location, while no match indicates the data block is in the original slow memory address, i.e., an identity mapping between the physical and device addresses. 
This metadata structure only needs to track the blocks in the fast memory, so its storage cost is moderate.

However, tag matching is viable only at low associativities. This is not only constrained by the hardware cost of associative search, but also due to the extra latency to access the tags from the memory. Usually the tag storage is too large to be kept on-chip, but they are accessed from off-chip memory. Assuming each tag is \SI{4}{B}, each 64-byte access can only retrieve 16 tags. Thus, for designs with associativities higher than 16, multiple metadata lookups are needed, incurring significant extra overheads.
\cref{fig:motiv:metadata} shows that although tag matching approaches the ideal efficiency at low associativities, its performance quickly drops at high-associativity cases.

An alternative approach involves a \emph{linear table} metadata format~\cite{pom14, mempod, chameleon, hybrid2}, a.k.a., a \emph{remap table} that tracks every data block in both the fast and slow memory tiers. Each access request uses the physical address to look up the remap table to find out the actual device address. Even at high associativities, the remap table lookup only needs a single memory access, more efficient than tag matching. Previous designs have proposed to use a simple on-chip \emph{remap cache} to further filter these off-chip remap table accesses~\cite{pom14, chameleon, mempod, hybrid2, atcache}. While a remap cache could achieve over 90\% hit rates, its scalability is limited considering the limited on-chip area vs. the increasing fast and slow memory capacities.

Furthermore, a more severe issue with linear remap tables is the storage cost. The remap table is stored in the fast memory, but it needs to cover all data blocks across both memory tiers and thus has a size growing with the slow memory capacity. At large slow-to-fast capacity ratios, the remap table would consume a substantial fraction of the fast memory, constraining the available space for actual data and hence degrading the performance. 
For instance, a \SI{16}{\GB} HBM paired with a \SI{512}{\GB} DRAM at the \SI{256}{B} granularity mandates over two billion entries. Assuming each entry is \SI{4}{B}, half of the fast memory would be spent on metadata storage.
Finer data block granularities, as proposed in certain studies~\cite{lh-cache, alloycache, cameo}, further enlarge the remap table size and exacerbate the storage overhead. 


In summary, the metadata \emph{lookup} and \emph{storage} overheads are becoming increasingly challenging and must be effectively addressed for future hybrid memory systems that have large slow-to-fast capacity ratios, high associativities, and finer block granularities. This is the focus of our work.


\section{Design}\label{sec:design}

We propose \design, a multi-level metadata structure for hybrid memory systems with efficient \emph{storage} and \emph{caching} approaches. \design is compatible with both cache and flat use modes of the fast memory, or even a hybrid of the two~\cite{optanedcpmm, chameleon, hybrid2, stealth-persist}. It is also completely transparent to application and system software. \design adopts an \emph{indirection-based remap table} (\irt) to effectively eliminate the need to store unnecessary metadata entries. It uses the saved fast memory space from the significantly reduced remap table size (up to 93\%) as an extra DRAM cache area to improve the performance. In addition, to overcome the extra latency overhead of looking up the multi-level table, \design also incorporates an efficient \emph{identity-mapping-aware remap cache} (\irc). By storing identity and non-identity address mappings with different cache entry formats, the overall hit rate of the remap cache increases, which further enhances the system performance.

\subsection{Design Overview}\label{sec:design:overview}

\begin{figure}
\centering
\includegraphics[width=\columnwidth]{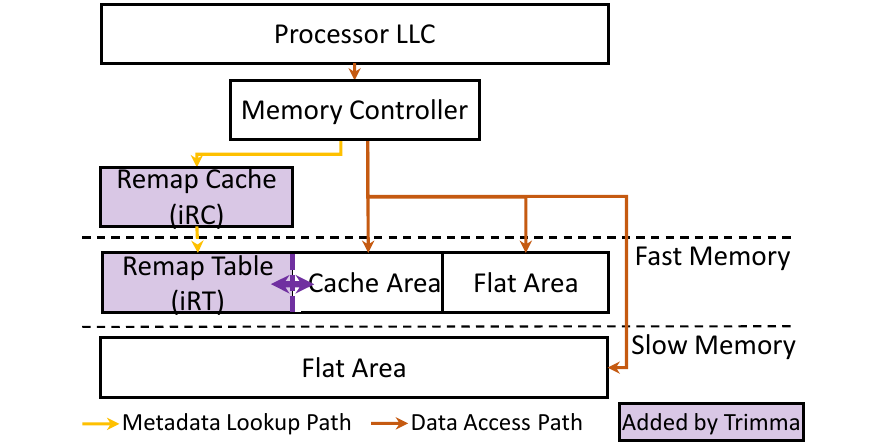}
\caption{The overview architecture of \design. Designs different from the baseline hybrid memory system are highlighted. 
The saved capacity from \irt can be flexibly used as extra cache space.}
\label{fig:overview}
\end{figure}

\begin{figure}
\centering
\includegraphics[width=\columnwidth]{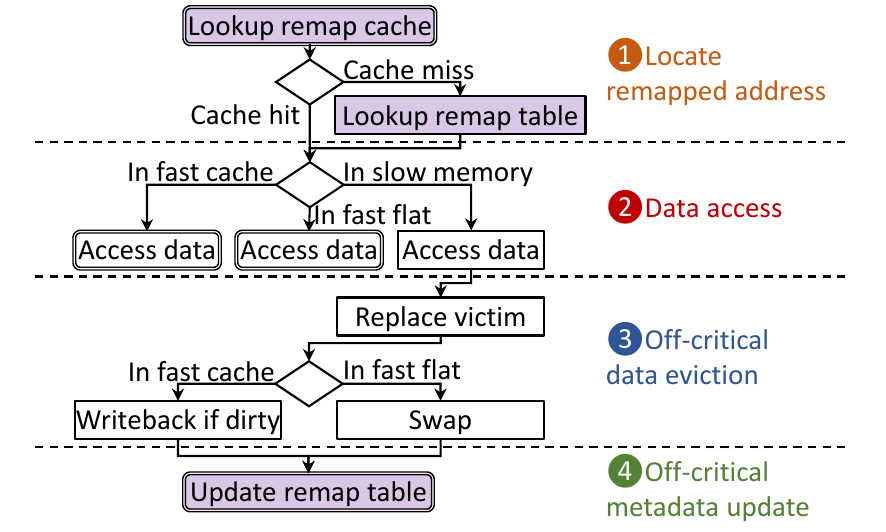}
\caption{The overall access flow of \design. Changes beyond the baseline hybrid memory system are highlighted.}
\label{fig:accessflow}
\end{figure}

\begin{figure}
\centering
\includegraphics[width=\columnwidth]{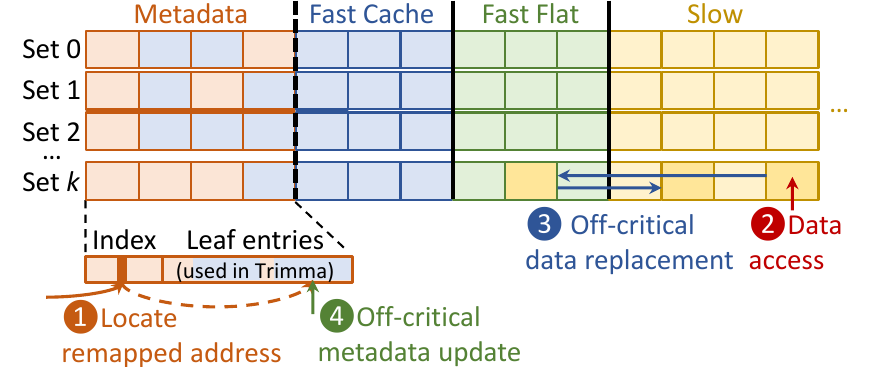}
\caption{The set-associative memory layout in \design, with the access flow actions in \cref{fig:accessflow} performed upon. \irt enables some unused metadata blocks to be used as extra cache space (shown in blue in the metadata area).}
\label{fig:memlayout}
\end{figure}

\definecolor{steponecolor}{RGB}{184, 96, 41}
\definecolor{steptwocolor}{RGB}{176, 36, 24}
\definecolor{stepthreecolor}{RGB}{56, 84, 146}
\definecolor{stepfourcolor}{RGB}{94, 129, 63}
\newcommand{\stepone}{\textcolor{steponecolor}{\ding{182}}}
\newcommand{\steptwo}{\textcolor{steptwocolor}{\ding{183}}}
\newcommand{\stepthree}{\textcolor{stepthreecolor}{\ding{184}}}
\newcommand{\stepfour}{\textcolor{stepfourcolor}{\ding{185}}}

\cref{fig:overview} illustrates the overall architecture of \design. \design is built on top of a basic hybrid memory design that uses the fast memory as either a DRAM cache or a part of the flat memory space, or even a combination of the two modes, such as Chameleon~\cite{chameleon} and Hybrid2~\cite{hybrid2}. Both the fast and slow memory spaces are divided into \emph{blocks}, as the granularity of caching/migration between the two tiers. 
To translate physical addresses to device addresses during data accesses, \design uses a forward-direction remap table as its metadata, similar to traditional linear remap tables but using a different data structure described later. This remap table is stored in the fast memory, and consumes its precious capacity which could have been used for the actual data.
All blocks in both memories are partitioned into disjoint \emph{sets} as a set-associative organization in \cref{fig:memlayout}. Blocks are cached/migrated between the two tiers only within each set, so each set tracks the metadata separately. \design is particularly efficient for high-associative configurations.

\cref{fig:accessflow} illustrates the overall access flow of accessing a physical address in the OS-visible flat area in either the fast or slow memory, and \cref{fig:memlayout} shows the corresponding actions on the memory layout. As data could be cached in or migrated to another place, we first determine the actual device address, by looking up the remap entry in the on-chip remap cache or from the remap table in the fast memory (\stepone). If the remap entry indicates that the data are in the fast memory (either the cache or flat area), we directly access it. Otherwise we fetch the data from the slow memory (\steptwo), return it to the processor, and handle replacement off the critical path (\stepthree\stepfour). The exact replacement policy choice is orthogonal to our design and discussed in \cref{sec:design:saving}. 

\cref{fig:accessflow} also highlights the changes in \design that adopt novel remap table and remap cache designs, which are merely in the metadata lookup and update phases (\stepone\stepfour) without affecting the rest data access/eviction.
Hence innovations including replacement policies~\cite{gran_aware_migration, hma15, thermostat, pageplacement-ics11, bw_aware_page_placement}, selective migration~\cite{banshee, batman, bear}, and cache-flat dual modes~\cite{chameleon, stealth-persist, memcached} can be orthogonally integrated with the remap table and remap cache designs in \design.

\textbf{Indirection-based remap table.}
Rather than the conventional linear remap table, \design uses an \emph{indirection-based remap table} (\irt, \cref{fig:multi_level}), inspired by the multi-level OS page tables in the x86-64 architecture. Essentially \irt resembles a generic radix tree structure, but is completely managed by hardware. Unnecessary remap entries in \irt are not allocated, in order to save the fast memory space used by metadata.
We identify and exploit two saving opportunities: (1) \emph{unallocated data blocks} are never accessed and do not need metadata~\cite{chameleon}; (2) \emph{data blocks that stay at their original locations without being cached/migrated} do not need to be translated, since they have identity address mappings (physical address == device address). While the first one is trivial and already leveraged by OS page tables, the second source is particularly effective and novel for hybrid memory systems. Specifically, with a high slow-to-fast capacity ratio, only a small subset of the slow memory blocks can be cached or migrated to the fast memory; the rest majority must stay in their original places due to the slow swap policy~\cite{pom14, silc-fm}. As a result, the total number of entries in \irt is proportional to only the \emph{fast memory capacity}, instead of the \emph{overall capacity}.

\irt is a much more scalable metadata design, as the slow memory capacity will be rapidly growing in the foreseeable future, while the fast memory size changes much slower due to the physical difficulties.
The saved fast memory spaces from the smaller \irt can be effectively utilized as extra caching slots that extend the existing ones, increasing the DRAM cache hit rate for higher performance.

Nevertheless, as a more complex structure than linear tables, \irt faces several design challenges, calling for unique optimizations that set it apart from OS page tables. First, as \irt is managed by hardware, its entry allocations/deallocations and updates must remain sufficiently simple and efficient while keeping the saving opportunities. Second, with frequent data block movements between the two memory tiers, the saved remap entries may scatter into a highly fragmented layout, and continuously come and go. It is hence quite difficult to effectively utilize these irregular spaces, and quickly reclaim them when needed by newly allocated metadata. Finally, we need to identify the optimized detailed configurations, including the number of levels and the tag bit width for each level. We address these issues in \cref{sec:design:irt,sec:design:saving}.

\textbf{Identity-mapping-aware remap cache.}
One other major concern of using a multi-level table is the increased lookup latency. An $L$-level remap table may introduce up to $L+1$ additional off-chip accesses in the worst case. 
Just like TLBs for page tables, a better caching approach is desired.

We observe that, we not only could skip storing an identity mapping in the remap table, we also \emph{do not need to look it up if we could know it is an identity mapping}, i.e., we simply use the original physical address as the device address to access the data. 
This motivates us to design our remap cache in an identity-mapping-aware manner, saving cache spaces by not storing redundant (identity) remapped addresses, and improving cache coverage and utilization.

However, a naive design of completely skipping identity mappings in the remap cache will not work. The key difficulty is that if we do not find an entry in the remap cache, we have \emph{no way to distinguish} between whether it is due to a miss to an actually valid remap entry, or because the entry is an identity mapping and thus skipped by the remap cache.
Therefore, if we think of a remap entry as a key (physical address) value (device address) pair, we still need to store the keys though the values can be saved.
As a result, we split the original single remap cache into two components in our \emph{identity-mapping-aware remap cache} (\irc): a \emph{NonIdCache} for valid remap entries as before, and an \emph{IdCache} to filter the skipped remap entries with better SRAM space utilization. Specifically, the IdCache uses a similar design as a sector cache~\cite{sectorcache}, and groups many entries into one cacheline to save space. Each entry only uses a single bit to indicate whether it is an identity mapping or not.
We discuss the detailed design in \cref{sec:design:irc}.

\subsection{Indirection-Based Remap Table}\label{sec:design:irt}

\begin{figure}
\centering
\includegraphics[width=\columnwidth]{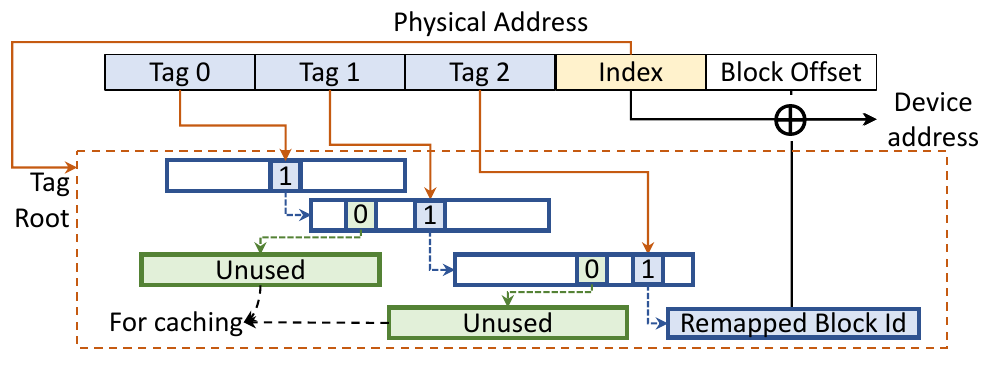}
\caption{The indirection-based remap table structure and its lookup flow.}
\label{fig:multi_level}
\end{figure}

\cref{fig:multi_level} illustrates the structure of \irt. To support set-associative hybrid memory systems, \irt uses separate trees for different sets. \irt is compatible with any associativity, and is particularly effective for high associativities when each set has more entries and thus more saving opportunities. 
Given a physical address, the index bits first select the \emph{tag root} for its set, which points to the root of the corresponding table. The tag bits in the physical address are divided into multiple parts, which are used to traverse the multi-level table, until locating a leaf entry that stores the remapped block ID to concatenate with the block offset to get the device address.

Different from OS page tables that are managed by kernel software, \irt is searched and updated purely in hardware. 
To simplify the management, we reserve a continuous space in the fast memory. The table of each set, which is a complete radix tree assuming all intermediate and leaf entries are allocated, is linearized to a standard linear layout in its breadth-first order. Because the tables of all sets are contiguously stored and have the equal allocated size, the tag root of a set can be directly located without using a lookup table. Furthermore, the address of any entry in a tree can also be inferred from the tag bits without being explicitly stored. Essentially, each entry, when present, always resides at its own reserved and fixed location. Thus the intermediate entries of \irt just use bit vectors to indicate whether the next-level entries are allocated or not, and only the leaf entries store the remapped block IDs.

\irt does not populate all entries in the table, and eliminates as many unnecessary entries as possible to reduce the table size in the fast memory. If the lookup misses in any level of the \irt, we assume an identity mapping with the device address equal to the physical address, as if the block has not been moved. Such a default also covers unallocated data blocks, which should be never accessed. However, the unused entries in \irt could be highly fragmented and scattered randomly in the fast memory. We describe how we effectively utilize them as cache spaces in \cref{sec:design:saving}.

\textbf{Example.}
\cref{fig:memlayout} illustrates an example access to a 2-level \irt. We perform one access to the intermediate level and the other to the leaf entry (\stepone). These accesses can happen in parallel as the entries are always in their fixed locations.
If the intermediate-level bit is 0 meaning that the leaf entry is unallocated (as in this example), the block is in the original location (\steptwo). After the accessed block is cached/migrated into the fast memory (\stepthree), we also need to update the remap table (\stepfour), by allocating the previously missing leaf entry and setting the corresponding upper-level bit (described below). This metadata block may have cached another data block, which must be evicted as described in \cref{sec:design:saving}.

\textbf{Table update.}
When a block is moved due to either caching, migration, eviction, or swap-out, its \irt entry should be accordingly updated. If the block is moved from its original location to another place, we allocate entries down to the leaf level; if it is restored back to its original slot, we clear the existing entry, and also free the upper levels if the entire block becomes empty. Such allocation/deallocation is simple because the \irt's linearized layout requires the entries to be at fixed locations. However, updates may involve changes in multiple levels. We reduce the cost by always buffering the intermediate-level entries in the on-chip controller when conducting the \irt lookup, so no backtracing is needed when these entries are later updated. 
In \design, a block will never be moved between two non-original locations; if it is evicted, it must go back to its initial place. This follows the slow swap policy~\cite{pom14, silc-fm}. 

\textbf{Detailed configuration selection.}
In \irt, we use 4-byte leaf entries to store the remapped block IDs. With a typical block size of \SI{256}{B}~\cite{hybrid2}, \irt can support up to $2^{32} \times \SI{256}{B} = \SI{1}{\TB}$ memory \emph{per set}. We can freely use multiple sets to support even higher capacity, e.g., using 1024 sets to cover \SI{1}{\peta B}. This is more than enough for hybrid main memories even considering future scaling.

Regarding the level division in \irt, more levels provide more metadata space saving opportunities, but also increase the overall lookup latency.
In order to conveniently utilize the saved metadata spaces without internal fragmentation, we require all \irt entries to be allocated/deallocated in a unit no smaller than the block size. This follows the same idea as the x86-64 page table which makes its intermediate levels always \SI{4}{\kB} aligned. 
With 256-byte blocks, each leaf metadata block in \irt would store 64 individual entries. Because the intermediate index levels store a single bit for each child instead of the address, each index block can hold 2048 children, corresponding to 11 bits for a tag chunk.
Larger granularities than a single block bring limited benefits (see \cref{sec:eval:sensitivity}), while requiring more complex multi-block eviction. So \design uses 11-bit tag chunks for all \irt levels.
This effectively realizes a 2048-ary radix tree. Because of the huge fanout, in this setting a simple 2-level \irt would be sufficient and more levels do not enable much additional space savings (\cref{sec:eval:sensitivity}). However, for even finer granularities such as \SI{64}{B} data blocks, deeper \irt{}s would be useful.

\textbf{Metadata storage savings.}
\irt saves significant metadata storage to be used as extra cache spaces. Assuming the above configurations of \SI{4}{B} remap entries, \SI{256}{B} blocks, and a 32:1 slow-to-fast capacity ratio (recall from \cref{sec:motivation:trends} that real-world systems may have a ratio up to 64:1~\cite{intel_spr}), a linear remap table occupies $(32+1) \times 4 / 256 = 52\%$ of the fast memory capacity, which further grows when the slow-to-fast ratio increases. If we apply a 2-level \irt design, the extra intermediate level has negligible storage overheads (worst-case $1/2048 = 0.05\%$) due to the use of valid bits rather than full addresses.
The number of valid leaf blocks depends on the specific workload. In the best case, all remapped entries (equal to the number of fast data blocks) are densely packed into a contiguous set of leaf metadata blocks, and we only consume $4 / 256 = 1.6\%$ of the fast memory, plus the intermediate level storage.
In the average cases, the leaf metadata blocks may be only partially occupied, and result in larger total occupation of metadata.
On average, \irt reduces the metadata size to 11.0\% of the fast memory.

\textbf{Novelty beyond OS page tables.}
While \irt shares a similar structure as OS page tables, it has several key unique novel features.
First, \irt is specially optimized for hardware, with predetermined and fixed addresses assigned to all the entries at different levels. This enables fast and parallel lookups, as well as efficient allocations/deallocations and updates, as described above.
Second, \irt exhibits a unique opportunity that avoids storing the metadata for uncached/unmigrated data blocks, which OS page tables are not able to do so. 
Finally, \irt also could effectively utilize the saved fast memory spaces as extra cache spaces to hold more data and improve performance. We describe it next.

\subsection{Using Saved Spaces for Caching}\label{sec:design:saving}

While \irt has a high potential to significantly reduce the metadata storage overhead and free up considerable fast memory capacity, it is not easy to effectively utilize the saved spaces. Available metadata blocks rapidly come and go when data blocks are cached/migrated/evicted, and they are highly scattered and fragmented in the fast memory with irregular distribution. Some prior hybrid memory designs such as Chameleon~\cite{chameleon} rely on the OS to manage the varying memory usage when it allocates/deallocates data. This approach will not work well when we additionally consider data blocks with identity mappings, whose states change with caching and migration, much more frequently than memory allocation. The software overheads would then become unacceptable.

Instead, in \design we keep the saved metadata spaces invisible to software, and use them as extra DRAM cache spaces managed completely in hardware. This allows us to adapt much faster to the quickly varying metadata size and timely exploit the short-term performance opportunities.
Notice that in order to cache a slow memory block into such an unused metadata slot, we use the same \irt to store the bidirectional mappings. The forward mapping (from the slow memory block to the DRAM cache block) is used for the look up process, which is illustrated in \cref{fig:multi_level}. The inverted mapping (from the DRAM cache block to the slow memory block) is used at eviction. In other words, to utilize one 256-byte unused block, we need to insert two 4-byte entries into the same \irt.

To track the availability of each metadata block and determine whether it can be used for caching, we reuse the \irt intermediate levels. For example, in a 2-level \irt, each leaf metadata block has a corresponding index bit, where ``1'' means it is used as metadata and ``0'' means it is unused. For a multi-level \irt, the availability of an intermediate-level block is recorded in its upper level. We describe how to use these bits for cache replacement below.

\textbf{Cache replacement of metadata and data.}
In order to simplify metadata allocation in hardware, the rigid \irt memory layout enforces each entry, if allocated, must reside in a certain fast memory block (\cref{sec:design:irt}). As a result, the metadata have higher priorities to use the block than the data. When an \irt update needs an entry, we directly evict the current data block cached in that location if there is one, regardless of its hotness. This data block could be refetched if needed and replace another less critical data block.

On the other hand, the replacement policy among data blocks should take into account the extra metadata slots if they are free to use. We distribute all reserved metadata blocks in the fast memory across all sets of the hybrid memory system, so each metadata block is dedicated to a certain set. If the corresponding index bit of this block is ``0'', it contributes to the set capacity. The difficulty is that the effective set associativity keeps varying over time, and also differs across different sets. We need to accordingly adapt the scope of data replacement policies.

Because \design is especially beneficial for highly associative systems, we extend popular replacement policies under high associativities to the extra cache space. 
Due to the high cost of tracking LRU information under high associativities, recent hybrid memory systems choose simple FIFO or random replacements~\cite{hybrid2, bimodal-cache, banshee}, or use area-efficient access counting~\cite{mempod}, in order to quickly locate an acceptable victim.
A FIFO policy simply skips blocks with index bits equal to ``1''. For a random policy, we check the index bit after randomly selecting a candidate, and resample if needed. Note that we can always evict a non-metadata block in the original cache area after a few times of retries. 

More complex policies, such as LRU~\cite{silc-fm, unisoncache}, CLOCK~\cite{clock_dwf}, and MEA in MemPod~\cite{mempod}, can also be applied. We provision enough replacement information storage for the maximum associativity including all metadata blocks.
This is a small overhead because the metadata area is usually smaller than or comparable to the basic cache area, and replacement information is a small portion compared to other data storage.
The hardware tracks the access behaviors for all blocks in the set as usual.
When an eviction is demanded, we use the index bits to skip actual metadata blocks, and find the best candidate in the remaining slots.

\design{} chooses to use the FIFO policy, which can further adopt a simple optimization to reduce the cost of off-chip index bit accesses when selecting victims. Specifically, due to the FIFO ordering, we can prefetch the next chunk of index bits to a small on-chip buffer (a few bytes would hide most latency).
We tried other more complex policies, e.g., an ideal LRU, but found only less than 1\% hit rate improvements (possibly because of the high associativity), while incurring huge off-chip update traffic~\cite{young2019update}.

\subsection{Identity-Mapping-Aware Remap Cache}\label{sec:design:irc}

The \irt in \design allows us to save the metadata storage, but exacerbates the metadata lookup latency and consumes more fast memory bandwidth by introducing multiple metadata accesses for each data block. Therefore we further propose an \emph{identity-mapping-aware remap cache} (\irc) to alleviate these issues.

\begin{figure}
\centering
\includegraphics[width=\columnwidth]{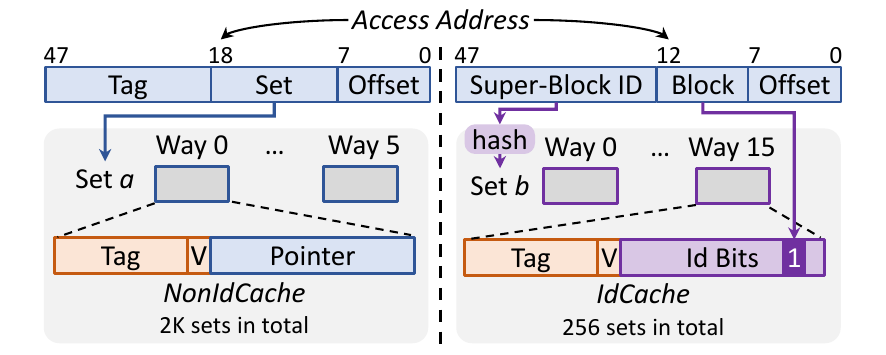}
\caption{The identity-mapping-aware remap cache design.}
\label{fig:irc_design}
\end{figure}

Under roughly the same SRAM capacity budget, we split the conventional remap cache into two components: a \emph{NonIdCache} that stores the valid (i.e., allocated) remap entries with non-identity mappings, and a special \emph{IdCache} that is organized similar to a sector cache~\cite{sectorcache} and stores bit vectors to indicate whether the entries associated with the data blocks in a larger \emph{super-block} use identity mappings or not.
Assuming the same SRAM capacity, the IdCache is able to store more entries due to the compressed format, thus improving the overall \irc coverage and runtime performance.

\cref{fig:irc_design} shows the cacheline formats of the NonIdCache and IdCache respectively. 
The NonIdCache operates in the same way as the conventional remap cache. In the IdCache, each cacheline uses the original space of a \SI{4}{B} remap pointer to store 32 bits that indicate whether the corresponding super-block of 32 contiguous data blocks (\SI{8}{\kB}) exhibits identity mappings or not. Each bit associates with one \SI{256}{B} block in this super-block.
Due to the different granularities (block vs. super-block) in the two parts, they use different index and tag schemes. For the IdCache, the lower address bits for the offset within the super-block are used to select one bit.
Moreover, we utilize a hash-based index scheme~\cite{kharbutli2004using} and a higher associativity in the IdCache to reduce conflict misses from the large number of identity mappings.
We empirically select the proper associativities to balance the access latencies of both caches. To summarize, the NonIdCache has 2048 sets of 6 ways each, and the IdCache has 256 sets of 16 ways each (\cref{tbl:configuration}). The total capacity is equal to a \SI{64}{\kB} conventional remap cache.

Each metadata data lookup would access the two caches in parallel, and either the IdCache or the NonIdCache would hit, or both miss. If hit in the IdCache, which means that a matched bit vector is found and the corresponding bit is 1, we directly use the original physical address as the device address. If hit in the NonIdCache, we use the pointer stored in the cache. If both are misses, we go to the off-chip remap table (i.e., \irt) to retrieve the entry. If the entry is valid, it is inserted into the NonIdCache. If the entry is not allocated, it is inserted into the IdCache.

Note that when \irt gets updated because of data block caching and migration, the corresponding \irc entries should also be updated for consistency. We simply invalidate the entries from \irc.

\textbf{Bloom filters as an alternative?}
At the first glance, the \irc could use Bloom filters~\cite{bloomfilter} as compressed storage. The physical block addresses with identity or non-identity mappings essentially form two sets, and the purpose of the \irc is to test whether a request address is in each set or not. 
However, due to the false positives in Bloom filters, we cannot use them to store the identity-mapping set; doing so may incorrectly classify an address with non-identity mapping into the identity-mapping set.
On the other hand, storing the non-identity-mapping set in Bloom filters has limited benefits; we still need to store their original remap entries in the cache.


\subsection{Discussion}\label{sec:design:dicuss}

\textbf{More saving opportunities.}
\design saves remap table entries for unallocated and uncached/unmigrated data blocks with identical physical-to-device address mappings, using a simple hardware-only solution. If we further leverage software support, there could be more opportunities. For example, when a data block is deallocated by the workload, it would never be accessed again and its remap entry can be recycled. Hardware alone cannot know this information, unless told by the software through a well-defined interface~\cite{chameleon}. We leave a detailed design as future work.

Another source of saving is to apply huge \emph{physical} pages, so that only one remap entry can be used for many data blocks. This is similar to huge OS pages that reduce virtual-to-physical mapping overheads~\cite{linux-thp, freebsd-thp, ingens, hawkeye, quicksilver}. While the current \irt structure naturally supports this, we find that without a co-designed OS to organize the physical page layout, the chances of contiguous data blocks together migrating between the fast and slow memories are quite low. We plan to explore this opportunity in the future.

\textbf{Fast swap policy.}
\design assumes the slow swap policy~\cite{pom14}, which provides more opportunities of identical mapping. Alternatively, a fast swap policy does not require an evicted block to go back to its initial location, so it avoids cascaded migration and potentially offers higher performance. In order to apply \irt to fast swap, we can adopt proactive migration in the background that exchanges slow memory blocks and restores their original locations, similar to Chameleon-Opt~\cite{chameleon}. However, this complicates the overall system and we do not further discuss it.

\section{Experimental Setup}\label{sec:methodology}

\textbf{System configurations.}
We use zsim~\cite{zsim}, a Pin-based simulator, to evaluate our designs. \cref{tbl:configuration} summarizes our detailed system configurations. 
We model two types of hybrid memory systems, HBM3 + DDR5, and DDR5 + NVM, both with a 32:1 slow-to-fast capacity ratio according to \cref{sec:motivation:trends}. The default block size is \SI{256}{B}. Other memory configurations are also considered in \cref{sec:eval:sensitivity}. The specifications of HBM3, DDR5, and NVM are extracted from recent literature and open-source implementations~\cite{ddr5, hbm3, ramulator, vans, pcm_energy}. We use CACTI~\cite{cacti} to model the SRAM-based \irc. 
We select the overall system capacity to be larger than the workload memory footprints (details below), so no application suffers from page faults. 

\begin{table}
  \centering\tablefontsize
  \caption{System configurations.}
  \label{tbl:configuration}
  \figvspace{-1em}
  \begin{tabular}{cl} 
    \toprule
    Cores       & x86-64, \SI{3.2}{\giga\hertz}, 16 cores    \\
    L1I         & \SI{32}{\kB} per core, 4-way, \SI{64}{B} cachelines, LRU   \\
    L1D         & \SI{64}{\kB} per core, 8-way, \SI{64}{B} cachelines, LRU   \\
    L2          & \SI{1}{\MB} per core, 8-way, 14-cycle latency, LRU    \\
    LLC         & \SI{32}{\MB} shared, 16-way, 60-cycle latency, LRU      \\
    \midrule
    \multirow{2}{*}{SRAM}
                & Conventional remap cache: 2048-set, 8-way, 3-cycle \\
                & \irc: 2048-set, 6-way (NonIdCache) + 256-set, 16-way (IdCache) \\
    \midrule
    \multirow{2}{*}{HBM3 + DDR5}
                & HBM 3.0, \SI{1600}{\mega\hertz}, 16 channels; RCD-CAS-RP: 48-48-48 \\
                & DDR5-4800, 1 channel, 2 ranks, 16 banks; RCD-CAS-RP: 40-40-40 \\
    \midrule
    \multirow{2}{*}{DDR5 + NVM}
                & DDR5-4800, 2 channels, 2 ranks, 16 banks; RCD-CAS-RP: 40-40-40 \\
                & NVM, \SI{1333}{\mega\hertz}, 2 channels, 1 rank, 8 banks; RD \SI{77}{\nano\second}, WR \SI{231}{\nano\second} \\
    \bottomrule
  \end{tabular}
\end{table}

\textbf{Workloads.}
We use the memory-intensive subset of SPECCPU 2017~\cite{spec2017} for multi-program workloads, similar to previous work~\cite{chameleon, silc-fm, pom14}. The others are insensitive to memory performance and \design has negligible impact on them.
For multi-threaded applications, we use the GAP benchmarks~\cite{gap}, the in-memory database silo~\cite{silo} with the TPC-C workload, and the memcached key-value store~\cite{memcached} with two workloads YCSB-A and YCSB-B~\cite{ycsb}. 
We run each SPEC benchmark in the rate mode with 16 copies, and use 16 threads for the multi-threaded workloads.
For SPEC, we fast-forward over the initialization phase and simulate 5 billion instructions. For each GAP workload, we manually mark and only simulate the 2nd to 5th iterations. 
For database workloads, we skip data loading and only simulate query execution. 
We use weighted speedup as the performance metric for multi-program workloads.

The original benchmarks have diverse memory footprints, ranging from \SI{6}{\GB} (16 processes of \texttt{519.lbm\_r}) to \SI{18}{\GB} (\texttt{twitter} in GAP). Thus we set the slow memory capacity to \SI{20}{\GB} to match the maximum footprint, and the fast memory to be 1/32 of it. Then, in the simulation we scale up each workload’s footprint to reach the full memory capacity, so roughly each workload can put 1/32 of its data in the fast memory. This ensures that our workloads touch large footprints and stress memory.
Note that although the total memory footprint exceeds the fast memory capacity, many applications still exhibit locality and an optimized high-associativity design is necessary to efficiently capture the active working set in the fast memory.


\textbf{Baselines.}
We evaluate our design under both the cache mode (\textbf{\design-C}) and the flat mode (\textbf{\design-F}) to show its general benefits. 
We compare \design-C with \textbf{Alloy Cache}~\cite{alloycache} and \textbf{Loh-Hill Cache}~\cite{lh-cache} as the cache mode baselines. Although Alloy Cache and Loh-Hill Cache utilize different formats to embed tags with the data or within the same DRAM row to optimize metadata accesses, they essentially adopt the tag matching strategy illustrated in \cref{sec:motivation} and are limited to low associativities. 
In Alloy Cache, metadata and data are accessed in a single burst, so we do not simulate extra metadata access cost. However, it is limited to the direct-mapped organization. 
For Loh-Hill Cache, we set the associativity to 30, i.e., 30 \SI{256}{B} blocks together with their metadata in one \SI{8}{\kB} DRAM row. We access metadata as DDR accesses with DRAM row buffer hits. Furthermore, we apply the RRIP replacement policy to Loh-Hill Cache, which offers a 2.1\% speedup over LRU.
In contrast, \design-C can scale to fully associative. We assume a perfect Memory Access Predictor in Alloy Cache and a perfect MissMap structure in Loh-Hill Cache, ignoring some of the metadata overheads to optimistically estimate their performance.
For flat mode designs, we compare \design-F with \textbf{MemPod}~\cite{mempod}. For both designs we use 4 sets with high associativities, as in MemPod. Both systems adopt the first-touch policy as in current NUMA systems~\cite{linux_numa_policy, numactl}, i.e., greedily allocating the workload data in the fast memory first, until its capacity is exhausted.

\section{Evaluation}\label{sec:eval}

\begin{figure*}
\centering
\includegraphics[width=\textwidth]{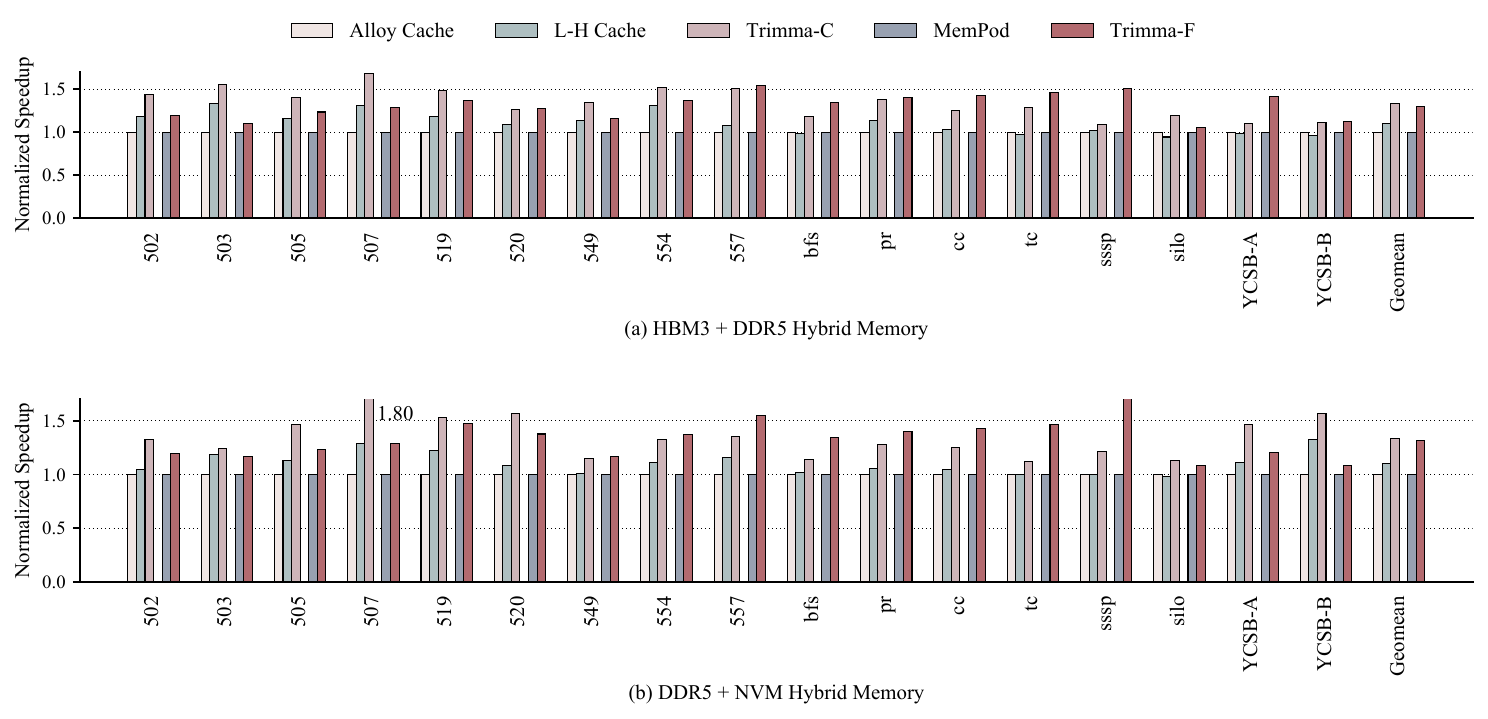}
\caption{Performance comparison between \design and the baselines on (a) HBM3 + DDR5 and (b) DDR5 + NVM. Performance of cache mode designs (Alloy Cache, L-H Cache, \design-C) is normalized to Alloy Cache. Performance of flat mode designs (MemPod, \design-F) is normalized to MemPod.}
\label{graph:speedup}
\end{figure*}

\subsection{Overall Performance Comparison}

\cref{graph:speedup} presents the overall performance of different designs under different memory technology combinations. All workloads we evaluate can benefit from the \design design. 
For the HBM3 + DDR5 system, on average, \design-C achieves 1.33$\times$ speedup over Alloy Cache while \design-F obtains 1.30$\times$ speedup over MemPod.
Because \design can provide more available fast memory spaces and scale to high associativities, it provides higher improvements for workloads with high memory footprint or high associativity requirements. Take \verb|557.xz| as an example. The low-associativity cache-mode baseline designs suffer from conflict misses in the fast memory and high migration traffic. In contrast, \design-C is 1.51$\times$ faster than Alloy Cache, by providing a higher associativity and more fast memory spaces. 
\design also excels if the workload exhibits more metadata savings, providing more extra caching spaces from the \irt design. This is the case for \verb|507.cactuBSSN_r|, where \irt reduces the metadata size by over 75\% and \design-C is 1.68$\times$ faster than Alloy Cache. 

A similar trend can also be observed with DDR5 + NVM. Overall \design-C achieves 1.34$\times$ speedup over Alloy Cache while \design-F obtains 1.32$\times$ speedup over MemPod. For workloads with higher memory footprints, e.g., \verb|sssp| with up to \SI{16}{\GB}, \design shows higher speedup on DDR5 + NVM, because the saved metadata space reduces the writeback traffic as in \cref{graph:hitrate_bw}(b), which significantly saves the limited slow memory bandwidth.

\begin{figure}
\centering
\includegraphics[width=\columnwidth]{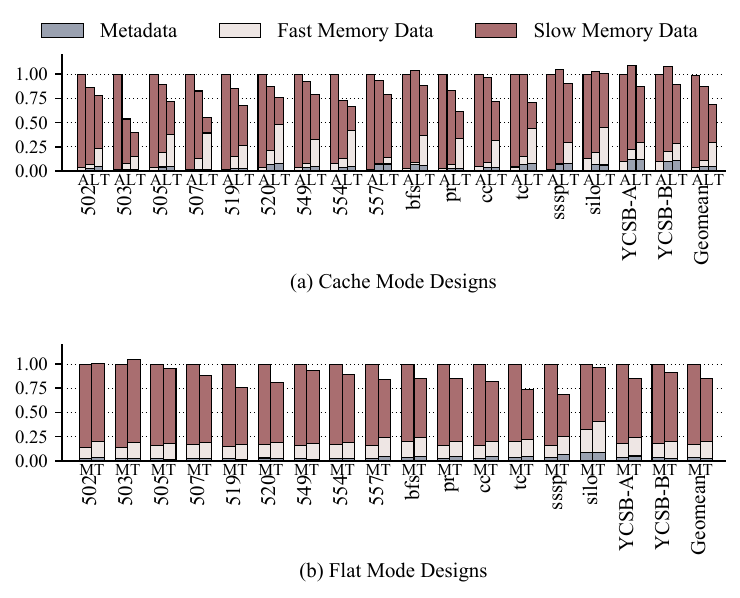}
\caption{Memory access latency breakdown on HBM3 + DDR5, for (a) the cache mode, including Alloy Cache ({A}), L-H Cache ({L}), and \design-C ({T}), and (b) the flat mode, including MemPod ({M}) and \design-F ({T}).}
\label{graph:memaccbreakdown}
\end{figure}

To further analyze the speedup sources, we break down the average memory access time into three parts: metadata lookup, fast memory data access, and slow memory data access, as shown in \cref{graph:memaccbreakdown}. 
Generally, \design-C effectively reduces the slow memory access time by over 50\% compared to Alloy Cache, where these accesses become hits in the fast memory, and increase the fast memory access time by 22\%. The metadata lookup cost is generally small over all designs, but \design-C increases this overhead by 4.6\%, due to more remap table accesses in \irt. Metadata lookups are insignificant because they are handled by the fast memory with low latency and high bandwidth, and many of them are filtered by the on-chip remap cache.


\subsection{Effectiveness Analysis of \irc and \irt}
\label{sec:eval:analysis}

\begin{figure}
\centering
\includegraphics[width=\columnwidth]{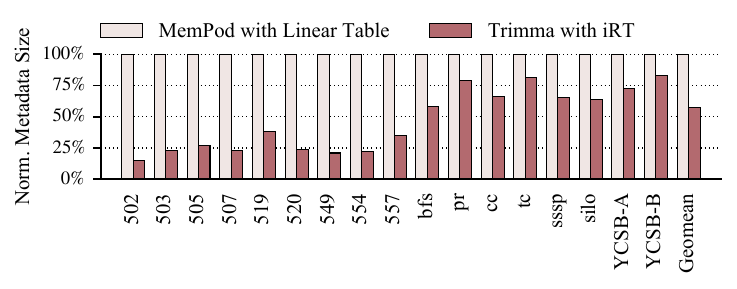}
\caption{Metadata size comparison between \design using \irt and MemPod using a linear table. The metadata size is obtained at the end of simulation.}
\label{graph:irt_saving}
\end{figure}

\begin{figure}
\centering
\includegraphics[width=\columnwidth]{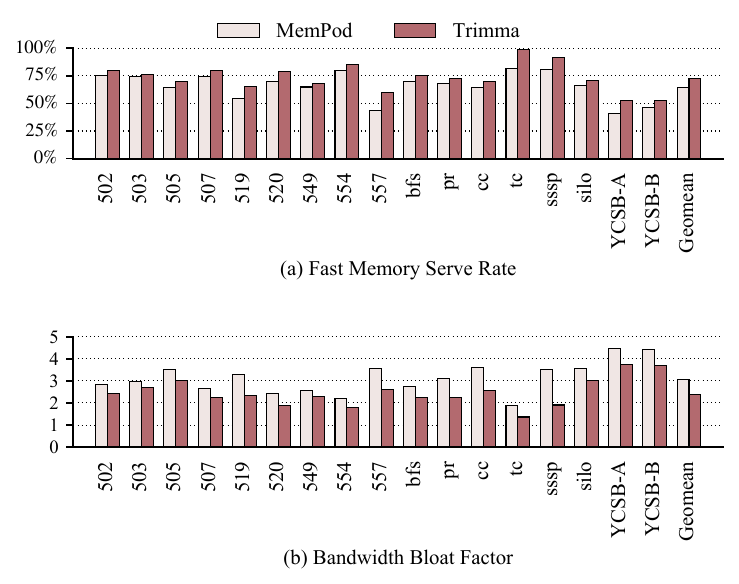}
\caption{
Detailed performance analysis between \design and MemPod, with (a) the percentage of memory accesses served by the fast memory (higher is better); (b) the ratio between total fast memory traffic and useful data traffic to the processor~\cite{bear} (lower is better).}
\label{graph:hitrate_bw}
\end{figure}

Next we use the flat mode designs (\design-F and MemPod) under the HBM3 + DDR5 configuration as a representative example, to explain the performance improvements from \irt and \irc. 

\textbf{\irt.}
\cref{graph:irt_saving} compares the metadata sizes of \irt and the baseline linear table. We capture the metadata sizes at the end of our simulation as it represents the most fragmented state of the memory.
\design \irt could effectively eliminate unnecessary identity mapping entries to save the metadata space by 43\% on average and up to 85\%. Higher spatial locality in the workload leads to higher savings. The saved spaces are used as extra caching spaces to improve performance. 
Similar to previous work~\cite{cameo, silc-fm}, we use the \emph{fast memory serve rate}, i.e., the percentage of accesses handled by the fast memory, and the \emph{bandwidth bloat factor}~\cite{bear}, i.e. the ratio between total fast memory traffic (including migration/swapping with the slow memory) and useful data traffic to the processor, to quantify the caching benefits and the migration overheads, respectively, as shown in \cref{graph:hitrate_bw}.
On average \design improves the fast memory serve rate by 7.9\%. Generally speaking, higher fast memory serve rate increases lead to more performance gains, e.g., \verb|519.lbm_r|, \verb|557.xz_r|, and \verb|tc|. Particularly, workloads with relatively low serve rates initially, e.g., \verb|557.xz_r|, have higher demands for fast memory spaces, which are exactly what \design provides by reducing the metadata space. 
On the other hand, \design also reduces 23\% memory migration traffic because of reduced conflict misses from larger available fast memory spaces. This is especially critical to the bandwidth-limited NVM-based slow memory.

On the other hand, \irt also incurs some overheads, as it requires multiple metadata accesses, rather than one access in a conventional linear table. However, these accesses in \irt are parallelized rather than serial (\cref{sec:design:irt}). Our experiments show that \irt only introduces at most 8.6\% extra latency cost (without a remap cache) and an average of 1\% (with \irc) compared to the single-probe case.
For updates, as in \cref{sec:design:irt}, all updates are buffered on-chip and written back to \irt together. This writeback occurs off the critical path and has negligible performance impact.

\begin{figure}
\centering
\includegraphics[width=\columnwidth]{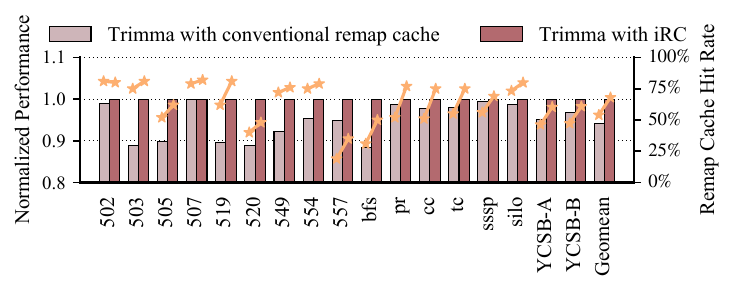}
\caption{Performance (bars) and remap cache hit rate (lines) comparison between the conventional remap cache and \irc.}
\label{graph:irc_benefit}
\end{figure}

\textbf{\irc.}
\cref{graph:irc_benefit} shows the benefits of our \irc design that separates the identity and non-identity mappings in the remap cache, compared to a conventional non-split remap cache. 
The overall remap cache hit rate increases from 54\% to 67\% on average, and the performance improves by 6.4\%.
We find that the conventional remap cache achieves good hit rates for non-identity mappings, while the hit rates are as low as 6\% for identity mappings. This is perhaps due to the relatively low hotness of identical mappings (otherwise these data blocks would be cached/migrated and become non-identity mappings). 
Using a dedicated \irc IdCache significantly increases the identity mapping hit rate to 32\%. Meanwhile, the non-identity part is not sensitive to the capacity loss because of the good locality of these hot entries~\cite{pom14, lgm}.

\subsection{Sensitivity Studies}
\label{sec:eval:sensitivity}

\begin{figure}
\centering
\includegraphics[width=\columnwidth]{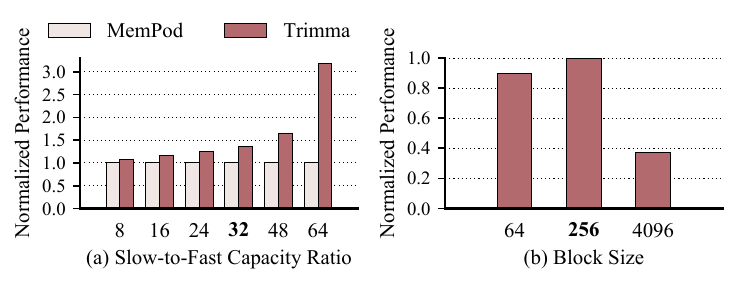}
\caption{Performance comparison between (a) different slow-to-fast capacity ratios (default is 32); and (b) different block sizes (default is \SI{256}{B}).
Performance is the geomean across all workloads.}
\label{graph:s2f_blk}
\end{figure}

\textbf{Slow-to-fast memory capacity ratios.}
\cref{graph:s2f_blk}(a) evaluates the speedups offered by \design under different capacity ratios between the slow and fast memories, from 8:1 to 64:1. Note that when the ratio reaches 64:1, the baseline linear remap table would occupy the entire fast memory capacity and all data are stored in the slow memory, rendering substantial performance degradation. On average, the speedup of \design increases roughly proportionally to the capacity ratio, varying from $1.07\times$ at 8:1 to $3.19\times$ at 64:1. Higher capacity ratios result in larger linear remap tables in the baseline, while the \irt size does not change much. Therefore \design enables more metadata space savings and thus allows more fast memory blocks to be used for application data.

\textbf{Block sizes.}
In \design we use \SI{256}{B} blocks as previous hybrid memory designs~\cite{hybrid2}. We also test other granularities and \cref{graph:s2f_blk}(b) shows the relative performance over \SI{256}{B}. 
The reason for low performance at \SI{64}{B} is that it fails to effectively exploit the spatial locality in the evaluated workloads and thus achieves lower hit rates. This result is validated by previous work~\cite{hybrid2}. When using the \SI{4}{\kB} granularity as the OS page size, it generally has similar or even higher hit rates compared to \SI{256}{B}, but the data over-fetching bandwidth consumption offsets the hit rate improvements, decreasing the overall performance by over 60\%.

\begin{figure}
\centering
\includegraphics[width=\columnwidth]{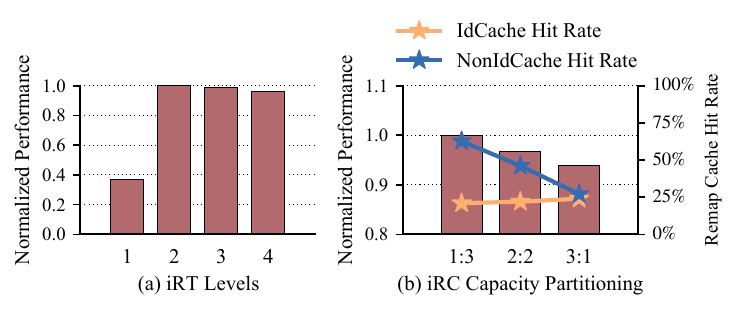}
\caption{Performance comparison between (a) different \irt schemes (default is 2-level \irt); and (b) different \irc capacity ratios between NonIdCache and IdCache (default is 1:3). Performance is the geomean across all workloads.
}
\label{graph:irt_irc}
\end{figure}

\textbf{\irt configurations.}
\cref{graph:irt_irc}(a) shows the performance with more multi-level schemes of \irt normalized to the default 2-level design in \design, including one level of bit vector and one level of the original remap table (organized in blocks). The single-level design simply falls back to the basic linear remap table, without the bit vector.
The four-level design is similar to Tag Tables~\cite{tagtables}, which slices the address tag part into four 6-bit chunks under \SI{256}{B} blocks. While Tag Tables did not use the saved space for caching, we exploit this opportunity in this comparison. Even so, its performance is worse than \design.
Although more levels provide more opportunities to use the intermediate level blocks for caching, we find that the increased metadata lookup cost and the additional complexity do not compensate for the metadata saving benefits, so the performance does not improve.

\textbf{\irc capacity partition.}
We evaluate several \irc capacity partitioning schemes and the results are shown in \cref{graph:irt_irc}(b). Although storing some identity mapping entries helps the overall hit rate and system performance, borrowing too much space from the non-identity-mapping cache would hurt. As mentioned in \cref{sec:eval:analysis}, identity mapping entries tend to be cold, so we do not need a large space to cache too many of them. Eventually we adopt 25\% space for identity mappings because it shows a high overall hit rate. 

\section{Related Work}\label{sec:related}

\textbf{Hybrid memory use modes.}
Flat-mode hybrid memory systems have larger OS-visible physical memory capacities to reduce page faults~\cite{pageplacement-ics11, hma15, pom14, cameo, silc-fm, mempod, lgm}, while cache-mode systems provide cheaper data migration for better performance~\cite{lh-cache, alloycache, unisoncache, footprintcache, footprint_tagless_cache, banshee, batman, atcache, dfc, bimodal-cache}. This tradeoff inspires researchers to consider combining both modes. Chameleon~\cite{chameleon} was the first design with flexible dual-mode support. It reused the unused space for caching, but required OS hints about data allocation. \design is instead software transparent and also exploits additional space saving opportunities.
Hybrid2~\cite{hybrid2} used a fixed caching space as a staging area for better migration. 
Baryon~\cite{baryon} supported data compression and sub-blocking for better capacity and bandwidth utilization. 
Stealth-persist~\cite{stealth-persist} proposed user-defined cache/flat area, but it focused on persistence rather than performance.

\textbf{Remap table design.}
Most hybrid memory systems managed by hardware use simple linear remap tables~\cite{pom14, mempod, chameleon, hybrid2}. Several DRAM cache designs~\cite{lh-cache, alloycache, unisoncache, bimodal-cache, enabling-fine-grain} make the metadata \emph{inlined} together with the data, e.g., in the same cacheline or DRAM row, so that a single access can fetch both. Such inlined metadata techniques are only applicable to direct-mapped or low-associativity systems, as prediction is required to decide a location before knowing the accurate metadata. Also, inlined metadata only hide the metadata lookup overhead but do not reduce their size. 

Another line of work eliminates the metadata cost by integrating the physical-to-device address translation with the OS virtual-to-physical address mapping~\cite{pageplacement-ics11, hma15, pageseer, thermostat, nimble, heteroos,  linux_tiering_proposal, kim2021exploring, hscc}. These designs require OS-level co-design and are less portable. Their performance is also limited in several aspects. First, the block size is restricted to coarse-grained \SI{4}{\kB} OS pages. Second, they only support epoch-based migration which cannot quickly adapt to program phase changes. Third, migration requires software involvement with significant interrupt and TLB shootdown overheads.


Tag Tables~\cite{tagtables} also adopted a multi-level remap table. \design is different from it. First, Tag Tables worked only for DRAM caches while \design supports both modes. Second, Tag Tables did not make use of the saved metadata space as \design does to improve performance. Third, \design simplifies metadata management by fixing their locations, while Tag Tables followed the complicated OS allocation. Nevertheless, many strategies in Tag Tables like lazy expansion and compressed entries can also be applied to \design.

\section{Conclusions}\label{sec:concl}

We propose \design, an indirection-based metadata structure and an efficient metadata cache design for hybrid main memory systems. \design integrates two techniques to address the metadata storage cost and lookup latency challenges. A multi-level remap table eliminates identical address mappings to save the metadata size in the fast memory. Performance is also improved when we utilize the saved space for extra data caching. We also propose a remap cache design that uses separate formats for non-identity and identity mapping entries, in order to increase the coverage and the hit rate. Overall, \design realizes scalable metadata management for future large-scale hybrid memory architectures.

\begin{acks}

The authors thank the anonymous reviewers and shepherd for their valuable suggestions, and the Tsinghua IDEAL group members for constructive discussion.
This work was supported by the National Natural Science Foundation of China (62072262). Mingyu Gao is the corresponding author.

\end{acks}

\balance
\bibliographystyle{ACM-Reference-Format}
\bibliography{refs}

\end{document}